\documentclass[aps,reprint,pra,longbibliography,superscriptaddress]{revtex4-1}
\usepackage{}
\usepackage{amssymb}
\usepackage{amsmath}
\usepackage{dcolumn}
\usepackage{bm}
\usepackage{graphicx}
\usepackage{mathrsfs}
\usepackage[colorlinks,linkcolor=blue,anchorcolor=blue,citecolor=blue,urlcolor=blue]{hyperref}
\begin{document}

\preprint{APS/123-QED}

\title{Phononic  waveguide assisted steady state entanglement of SiV centers}

\author{Yi-Fan Qiao}
\affiliation{Shaanxi Province Key Laboratory of Quantum Information and Quantum Optoelectronic Devices,
Department of Applied Physics, Xi'an Jiaotong University, Xi'an 710049, China}
\author{Hong-Zhen Li}
\affiliation{Shaanxi Province Key Laboratory of Quantum Information and Quantum Optoelectronic Devices,
Department of Applied Physics, Xi'an Jiaotong University, Xi'an 710049, China}
\author{Xing-Liang Dong}
\affiliation{Shaanxi Province Key Laboratory of Quantum Information and Quantum Optoelectronic Devices,
Department of Applied Physics, Xi'an Jiaotong University, Xi'an 710049, China}
\author{Jia-Qiang Chen}
\affiliation{Shaanxi Province Key Laboratory of Quantum Information and Quantum Optoelectronic Devices,
Department of Applied Physics, Xi'an Jiaotong University, Xi'an 710049, China}
\author{Yuan Zhou}
\affiliation{ School of Science, Hubei University of Automotive Technology, Shiyan 442002, China}
\author{Peng-Bo Li}
 \email{lipengbo@mail.xjtu.edu.cn}
\affiliation{Shaanxi Province Key Laboratory of Quantum Information and Quantum Optoelectronic Devices,
Department of Applied Physics, Xi'an Jiaotong University, Xi'an 710049, China}

\date{\today}% It is always \today, today,

\begin{abstract}
Multiparticle entanglement is of great significance for quantum metrology and quantum information processing.
We here present an efficient scheme to generate stable multiparticle entanglement in a solid state setup, where an array of silicon-vacancy centers are embedded in a quasi-one-dimensional  acoustic diamond waveguide.
In this scheme, the continuum of phonon modes induces a controllable dissipative coupling among the SiV centers.
We show that, by an appropriate choice of the distance between the SiV centers, the dipole-dipole interactions can be switched off due to destructive interferences, thus realizing a Dicke superradiance model. This gives rise to an entangled steady state of SiV centers with high fidelities.
The protocol provides a feasible setup for the generation of multiparticle entanglement in a solid state system.

%\begin{description}
%42.50.Dv, 03.67.-a, 76.30.Mi
%\end{description}
\end{abstract}

\maketitle

%\tableofcontents

\section{introduction}
Multiparticle entanglement has attracted  great attention for its diverse applications in quantum metrology and quantum computing \cite{PhysRevA.67.013607,PhysRevLett.123.073001,PhysRevLett.110.080502,Liu_2015,Li:11,PhysRevA.93.012341,PhysRevLett.88.197901,S_rensen2001,PhysRevA.59.2468,PhysRevA.79.042334}.
Efforts along this direction have lead to a plenty of proposals for generating entangled states with particles as many as possible \cite{Wang2018,PhysRevLett.117.210502,PhysRevLett.119.180511}, such as spin squeezing states \cite{PhysRevB.94.205118,PhysRevA.95.040301,PhysRevLett.107.206806} and GHZ states \cite{PhysRevLett.106.110402,PhysRevA.100.052302,Li:20,PhysRevA.101.012345}.
So far, multiparticle entanglement has been realized involving up to 20 qubits in trapped-ion systems \cite{PhysRevX.8.021012}, and 12 qubits in superconducting circuits \cite{PhysRevLett.122.110501}.
A dominating challenge to generate high-quality entangled states in the experiment comes from environment noises \cite{Luo620}.
The particles inevitably interact with the environment,  thus inducing  the decoherence effect \cite{Verstraete2009}.
As a result, these entangled states are extremely fragile and easily destroyed.
A feasible approach to solve this problem is using dissipation as a resource.
Coupling to the environment drives the system to a steady state.
By taking advantage of the dissipation, one can engineer a large variety of strongly correlated states in steady state \cite{Song:17,Stannigel_2012,PhysRevA.88.013837,PhysRevA.57.548,PhysRevA.83.052312,PhysRevLett.106.090502,PhysRevA.100.012339,PhysRevLett.110.120402,PhysRevA.100.052332,PhysRevLett.107.080503,PhysRevLett.108.043602,PhysRevA.90.054302}.

Color centers in diamond, such as germanium-vacancy (GeV) \cite{PhysRevLett.118.223603}, nitrogen-vacancy (NV) center \cite{PhysRevA.98.052346,Childress281,DOHERTY20131,PhysRevApplied.10.024011,PhysRevApplied.11.044026} and silicon-vacancy (SiV) center \cite{PhysRevLett.113.263601,Becker_2016,PhysRevLett.122.063601,Lemonde_2019}, play an important role in quantum science and technology \cite{doi:10.1002/qute.201900069,PhysRevA.83.054306,Casola2018,PhysRevLett.105.210501,PhysRevApplied.4.044003}.
Due to the high controllability and long coherence time \cite{Awschalom1174,PhysRevA.97.052303,PhysRevX.6.041060,Maze_2011,PhysRevLett.108.043604,PhysRevLett.108.143601,PhysRevLett.117.015502,PhysRevA.96.063810,Balasubramanian2009}, NV centers stand out among all kinds of solid-state systems.  However, despite the impressive achievement with these solid-state systems, it is still challenging to realize quantum  information processing with a large number of spins,
due to the inherent weak coupling of NV spins to phonon modes and  the difficulty in scaling to many spins.
Recently, much attention has in particular been paid to the study of SiV centers \cite{PhysRevApplied.11.024073,PhysRevLett.107.235502,Lekavicius:19}.
The negatively charged SiV centers present strong zero-phonon line emission and narrow inhomogeneous broadening, which result from its inversion symmetry \cite{PhysRevLett.113.263602,PhysRevLett.112.036405,PhysRevB.88.235205}.
These characters make SiV centers become an excellent candidate as quantum units, due to their favorable optical properties and flexible manipulation on the energy-level structures \cite{PhysRevLett.119.223602,Neu_2013,PhysRevApplied.5.044010}.
Moreover, it has been discussed that SiV centers can realize strong and tunable coupling to phonons at the single quantum level \cite{PhysRevB.97.205444,PhysRevB.94.214115,Sohn_2018,PhysRevLett.120.213603,PhysRevResearch.2.013121}.

In this paper, we propose an efficient scheme to generate steady state entanglement of solid-state qubits with an array of SiV centers placed in a one dimensional (1D)  phononic diamond waveguide.
A crucial advantage for utilizing the phononic waveguide is that SiV centers can be positioned at determined positions to modulate the phase difference \cite{PhysRevA.68.012101,PhysRevLett.101.113903}.
In addition, the electronic ground states of  SiV centers feature orbital degrees of freedom, which permits a direct and strong strain coupling to the phonon modes. Tailoring the system dissipation via the auxiliary external driven fields, the separated SiV centers are collectively coupled to the propagating phonon modes. Thus, we can engineer these SiV spins to a steady entangled state.
By suitably choosing the distances of SiV centers, the dipole-dipole interactions will vanish due to destructive interferences \cite{PhysRevB.82.075427,Chang_2012,PhysRevA.95.033818}, and we can achieve a dissipative Dicke supperradiant \cite{PhysRevLett.112.140402} model in this setup \cite{PhysRevA.99.023822,PhysRev.93.99}.
This scheme provides a promising avenue for the generation of  many-body entanglement at the steady state  in solid-state setups.

\begin{figure}
\includegraphics[width=8cm]{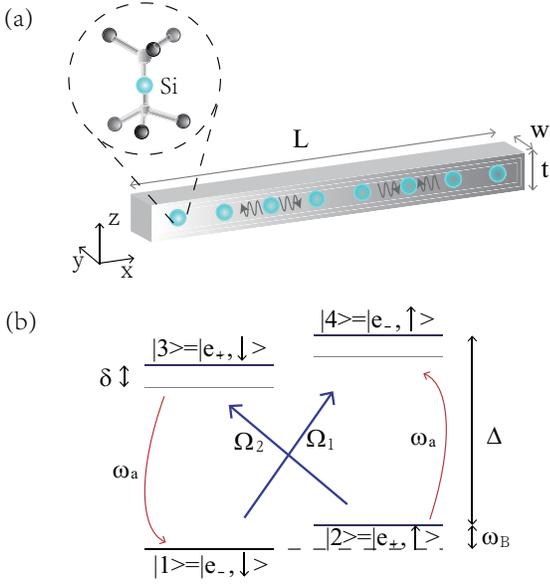}
\caption{\label{fig:1}(Color online)
(a) Sketch of an array of $N$ SiV centers are embedded in a 1D phononic diamond waveguide at fixed positions $x_{j}$,
with the uniform distance between two nearby SiV centers.
The length, width and thickness of the waveguide are $L$, $w$, and $t$, respectively.
The orbital states of the centers are coupled via strain to a continuum of compression modes propagating along the waveguide (indicate by the curvy arrow).
(b) Ground state of SiV centers.
Two time-dependent microwave driving fields induce the Raman processes between $|3\rangle\leftrightarrow|2\rangle$, and $|4\rangle\leftrightarrow|1\rangle$. }
\end{figure}

\section{Model}
As depicted in Fig.~\ref{fig:1}(a),  an array of $N$ SiV centers are coupled to the phonon modes in a 1D diamond waveguide.
The lattice distortion of longitudinal compression modes affects the electronic structure of the defect, which gives rise to a strain coupling between the phonons and the orbital degrees of freedom of the SiV centers \cite{PhysRevB.97.205444,PhysRevLett.120.213603}.
The Hamiltonian for the whole system is given by
\begin{eqnarray}\label{ME1}
\hat{H}&=&\hat{H}_{\text{SiV}}+\hat{H}_{\text{ph}}+\hat{H}_{\text{strain}}.
\end{eqnarray}
The first term $\hat{H}_{\text{SiV}}$ is the Hamiltonian of the $N$ SiV centers. The second term $\hat{H}_{\text{ph}}$ corresponds to the
quantized Hamiltonian of  the phononic waveguide modes. The last term $\hat{H}_{\text{strain}}$ describes the strain coupling between the
SiV centers and the phonon modes.

The SiV center in diamond is formed by a silicon atom and a split vacancy replacing two neighboring carbon atoms [see Fig.~\ref{fig:1}(a)] \cite{PhysRevLett.112.036405}.  The Hamiltonian of a single SiV center includes three interaction parts: the spin-orbit coupling $\hat{H}_{\text{SO}}$, the Jahn-Teller interaction $\hat{H}_{\text{JT}}$, and the Zeeman interaction $\hat{H}_{\text{Z}}$.
As the $L_{x}$ and $L_{y}$ components of the angular momentum $\vec{L}$ are zero in the basis spanned by the degenerate eigenstates $|e_{x},\uparrow\rangle$, $|e_{x},\downarrow\rangle$, $|e_{y},\uparrow\rangle$ and $|e_{y},\downarrow\rangle$, the spin-orbit interaction can be simplified to $\hat{H}_{\text{SO}}=-\lambda_{g}L_{z}S_{z}$, where $\lambda_{g}=2\pi\times45$ GHz is the spin-orbit coupling \cite{PhysRevLett.112.036405}. In the presence of external magnetic fields, the Zeeman term is $\hat{H}_{\text{Z}}= f \gamma_{L} L_{z}B_{z}+\gamma_{S}\vec{S}\cdot\vec{B}$, where $\gamma_{L} $ and $\gamma_{S} $ are the orbital and spin gyromagnetic ratios, respectively. The quenching factor $f\approx0.1$ is caused by the Jahn-Teller coupling \cite{PhysRevLett.112.036405}.
Assuming $\vec{B}=B_{0}\vec{e}_{z}$, we have the following Hamiltonian with the matrix form
\begin{eqnarray}\label{ME2}
\hat{H}_\text{SiV}&=&-\lambda_{g}\begin{bmatrix} 0 &i \\ -i &0 \end {bmatrix} \otimes \frac{1}{2} \begin{bmatrix} 1&0 \\ 0&-1 \end {bmatrix}+\begin{bmatrix} \Upsilon_{x}&\Upsilon_{y} \\ \Upsilon_{y}&-\Upsilon_{x} \end {bmatrix} \otimes \mathbb{I} \notag\\
&+&f\gamma_{L} \begin{bmatrix} 0 &i \\ -i &0 \end {bmatrix} \otimes B_{0}\mathbb{I}+\gamma_{S}B_{0} \mathbb{I}  \otimes \frac{1}{2}\begin{bmatrix} 1 &0 \\ 0 & -1 \end {bmatrix}.
\end{eqnarray}
Here, $\Upsilon_{x}$ and $\Upsilon_{y}$ are the distortions along $x$ and $y$, and $\mathbb{I}$ denotes the $2\times2$ identity matrix.
When neglecting the effect of the reduced orbital Zeeman interaction, the third term in Eq.~(\ref{ME2}) can be ignored \cite{PhysRevLett.120.213603}.
Diagonalizing the above equation and in the case of $\Upsilon_{x,y}\ll\lambda_{g}$, we obtain the two lower eigenstates $|1\rangle\approx|e_{-},\downarrow\rangle$, $|2\rangle\approx|e_{+},\uparrow\rangle$, and two upper eigenstates $|3\rangle\approx|e_{+},\downarrow\rangle$, $|4\rangle\approx|e_{-},\uparrow\rangle$, with the splitting given by $\Delta=\sqrt{\lambda_{g}^{2}+4(\Upsilon_{x}^{2}+\Upsilon_{y}^{2})}\approx2\pi\times46$ GHz \cite{PhysRevLett.120.213603}. Here, $|e_{\pm}\rangle=(|e_{x}\rangle\pm i|e_{y}\rangle)/\sqrt{2}$ are the eigenstates of the angular momentum $L_{z}$.
We consider adding  two time-dependent driving fields to induce the transition between  $|1\rangle $ and $|4\rangle$, with the amplitude $\Omega_{1}$ and frequency $\omega_{1}$, and the transition between  $|2\rangle $ and $|3\rangle$, with the amplitude $\Omega_{2}$ and frequency $\omega_{2}$.
Thus, the Hamiltonian of a single SiV center driven by external fields can be described by
\begin{eqnarray}\label{ME3}
\hat{H}_{\text{SiV}}&=&\omega_{B}|2\rangle\langle2|+\Delta|3\rangle\langle3|+(\Delta+\omega_{B})|4\rangle\langle4| \notag\\
&+&\frac{\Omega_{1}}{2}|1\rangle\langle4|e^{i\omega_{1}t}+\frac{\Omega_{2}}{2}|2\rangle\langle3|e^{i\omega_{2}t}+\text{H.c.},
\end{eqnarray}
where $\omega_{B}=\gamma_{s}B_{0}$ is the Zeeman energy.
The energy level scheme and related transitions associated with $\hat{H}_{\text{SiV}}$ are summarized in Fig.~\ref{fig:1}(b).

\begin{figure}
\includegraphics[width=6cm]{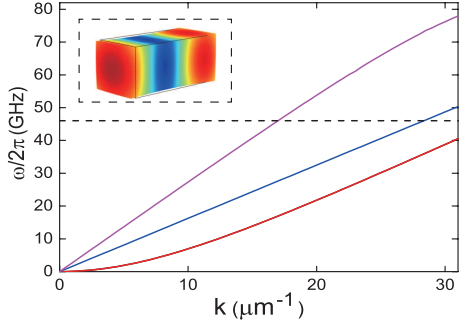}
\caption{\label{fig:2}(Color online)
 Acoustic dispersion relation for a rectangular waveguide of width and thickness $w=t=100$ nm.
 Inset: Normalized displacement profiles of symmetric phonons at 46 GHz.}
\end{figure}

We now consider the Hamiltonian for the phonon modes in the diamond waveguide. For 1D diamond phononic waveguide, the length, width and thickness are labelled as $L$, $w$ and $t$, which satisfy the condition $L\gg wt$.
For a linear isotropic medium, we can model the phonon modes as elastic waves with a displacement field $\vec{u}(\vec{r},t)$ \cite{PhysRevLett.120.213603}.
Imposing the periodic boundary conditions and then quantizing the displacement field, we can obtain the quantized Hamiltonian \cite{Stroscio_1996} of the phonon modes
\begin{eqnarray}\label{ME4}
\hat{H}_{\text{ph}}=\sum_{n,k}\hbar\omega_{n,k}a_{n,k}^{\dag}a_{n,k}.
\end{eqnarray}
Here, $\omega_{n,k}$ is the frequency of phonon modes, with $k$  the wave vector along the waveguide and $n$  the branch index, and $a_{n,k}$ is the bosonic annihilation operator for the phonon mode.
In addition, the quantized displacement field reads
\begin{eqnarray}\label{ME5}
\vec{u}(\vec{r})=\sum_{n,k}\sqrt{\frac{\hbar}{2\rho V\omega_{n,k}}}\vec{u}_{n,k}^{\perp}(y,z)(a_{n,k}e^{ikx}+\text{H.c.}),
\end{eqnarray}
where $\rho$ is the density, $V=Lwt$ is the volume of the waveguide, and $\vec{u}_{n,k}^{\perp}(y,z)$ is the transverse profile of the displacement field \cite{PhysRevLett.120.213603}.
In Fig.~\ref{fig:2}, we summarize the simulated acoustic dispersion relation and the displacement distribution for the waveguide with $w=t=100$ nm.

We proceed to discuss the strain coupling between the SiV centers and the phononic waveguide modes. The change in Coulomb energy of electronic states due to a collective displacement of the defect atoms will induce strain coupling between these phonons and the orbital degrees of freedom of SiV centers \cite{PhysRevB.94.214115}. When considering the small displacement of the defect atoms, the coupling between the phonon modes and the orbital degrees of freedom is linear in the Born-Oppenheimer approximation \cite{PhysRevB.97.205444,PhysRevLett.120.213603,PhysRevB.94.214115}. In this case, the strain coupling within the framework of linear elasticity theory can be given by
\begin{eqnarray}\label{ME6}
\hat{H}_{\text{strain}}=\epsilon_{E_{gx}}(\hat{L}_{-}+\hat{L}_{+})-i\epsilon_{E_{gy}}(\hat{L}_{-}-\hat{L}_{+}),
\end{eqnarray}
where $\hat{L}_{+}=\hat{L}_{-}^{\dag}=|3\rangle\langle1|+|2\rangle\langle4|$ is the orbital raising operator within the ground state.
The parameters $\epsilon_{E_{gx}}$, and $\epsilon_{E_{gy}}$ are
\begin{eqnarray}\label{ME7}
\epsilon_{E_{gx}}&=&d(\epsilon_{xx}-\epsilon_{yy})+f\epsilon_{zx} \notag\\
\epsilon_{E_{gy}}&=&-2d\epsilon_{xy}+f\epsilon_{yz},
\end{eqnarray}
with $\epsilon_{ab}$   the strain field tensor \cite{PhysRevB.94.214115}.

After taking Eq.~(\ref{ME5}) and Eq.~(\ref{ME7}) into Eq.~(\ref{ME6}), we obtain the resulting strain coupling for $N$ SiV centers
\begin{eqnarray}\label{ME8}
\hat{H}_{\text{strain}}=\sum_{j,n,k}[g_{n,k}^{j}(J_{+}^j+J_{-}^j)a_{n,k}e^{ikx_{j}}+\text{H.c.}].
\end{eqnarray}
Here $J_{-}^j=(J_{+}^j)^{\dag}=|1\rangle_j\langle3|+|2\rangle_j\langle4|$ is the spin-conserving lowering operator and $j$ labels the SiV center located at the position $x_{j}$.
In order to get the right expression for the dipole-dipole force in the next part, it is crucial to keep the counter-rotating terms in the Hamiltonian of strain coupling \cite{PhysRevLett.110.080502}. After some derivation, the resulting coupling strength can be given as
\begin{eqnarray}\label{ME9}
g_{n,k}^{j}=d\sqrt{\frac{\hbar k^{2}}{2\rho V\omega_{n,k}}}\xi_{n,k}(y_{j},z_{j}),
\end{eqnarray}
where $d/2\pi\sim1$ PHz is the strain sensitivity \cite{Sohn_2018,PhysRevLett.120.213603}. Here, the dimensionless function $\xi_{n,k}(y,z)$ is related to the specific strain distribution.
Putting everything together, the total Hamiltonian of the whole system  is given by
\begin{eqnarray}\label{ME10}
\hat{H}_{\text{Total}}&=&\sum_{j}[\omega_{B}|2\rangle_{j}\langle2|+\Delta|3\rangle_{j}\langle3|+(\Delta+\omega_{B})|4\rangle_{j}\langle4| \notag\\
&+&\frac{\Omega_{1}}{2}(|1\rangle_j\langle4|e^{i\omega_{1}t}+\text{H.c.})+\frac{\Omega_{2}}{2}(|2\rangle_j\langle3|e^{i\omega_{2}t}+\text{H.c.})] \notag\\
&+&\sum_{n,k}\omega_{n,k}a_{n,k}^{\dag}a_{n,k} \notag\\
&+&\sum_{j,n,k}[g_{n,k}^{j}(J_{+}^j+J_{-}^j)a_{n,k}e^{ikx_{j}}+\text{H.c.}].
\notag\\
\end{eqnarray}

\section{Steady state entanglement}
We use the quantum theory of damping in which the waveguide modes are treated as a reservoir.
In this case, the phonon modes can be eliminated by using the Born-Markov approximation, giving rise to  an effective master equation for the density operator $\hat{\rho}$, which describes the dissipative dynamics of the spin degrees of freedom.
Starting from the model given in Eq.(\ref{ME10}), we make a rotating transformation $H\rightarrow UHU^{\dag}+iU^{\dag}\dot{U}$ firstly, where the unitary operation $U=e^{i(\omega_{1}|4\rangle\langle4|+\omega_{2}|3\rangle\langle3|)t}$ \cite{PhysRevLett.110.080502}.
The resulting Hamiltonian in the case of a single SiV center at the position $x_{j}=0$ is
\begin{eqnarray}\label{ME11}
\hat{H}&=&\sum_{n,k}\omega_{n,k}a_{n,k}^{\dag}a_{n,k}+\omega_{B}|2\rangle\langle2|+\Delta_{2}|3\rangle\langle3|+\Delta_{1}|4\rangle\langle4| \notag\\
&+&\frac{\Omega_{1}}{2}(|1\rangle\langle4|+\text{H.c.})+\frac{\Omega_{2}}{2}(|2\rangle\langle3|+\text{H.c.}) \notag\\
&+&\sum_{n,k}g_{n,k}[(|3\rangle\langle1|e^{i\omega_{2}t}+|4\rangle\langle2|e^{i\omega_{1}t}+\text{H.c.})a_{n,k}+\text{H.c.}],
\notag\\
\end{eqnarray}
where $\Delta_{1}=\Delta+\omega_{B}-\omega_{1}$ and $\Delta_{2}=\Delta-\omega_{2}$.
A crucial step to get the effective Hamiltonian is the adiabatic elimination of the excited states.
Next, we apply the following canonical transformation $H\rightarrow e^{-S}H e^{S}$ \cite{PhysRevLett.110.080502,PhysRev.149.491}, where
\begin{eqnarray}\label{ME12}
S=\frac{\Omega_{1}}{2\Delta_{1}}(|4\rangle\langle1|-|1\rangle\langle4|)+\frac{\Omega_{2}}{2\Delta_{2}}(|3\rangle\langle2|-|2\rangle\langle3|).
\end{eqnarray}
Note that this transformation is valid in the limit of $\Omega_{1,2}\ll\Delta_{1,2}$.
Keeping terms to the second order, and neglecting the terms which are proportional to the excited state populations, we obtain the effective Hamiltonian in the interaction picture with respect to $H_{0}=\sum_{n,k}\omega_{n,k}a^{\dag}_{n,k}a_{n,k}+\omega_{B}|2\rangle\langle2|$
\begin{eqnarray}\label{ME13}
\hat{H}_{\text{I}}&=&\alpha u(x_{0},t)(D^{+}e^{i\omega_{a}t}+\text{H.c.}),
\end{eqnarray}
with $u(x_{0},t)=\sum_{n,k}g_{n,k}(a_{n,k}e^{-i\omega_{a}t}+\text{H.c.})$.  $D^{-}=(D^{+})^{\dag}=u|1\rangle\langle2|+\upsilon|2\rangle\langle1|$ is a jump operator resulting from the cross radiative decay, with $u=\alpha^{-1}\Omega_{2}/2\Delta_{2}$ and $\upsilon=\alpha^{-1}\Omega_{1}/2\Delta_{1}$ satisfying the relation $u^{2}-\upsilon^{2}=1$, and $\alpha^{2}=(\Omega_{1}/2\Delta_{1})^{2}-(\Omega_{2}/2\Delta_{2})^{2}$ is a normalization constant.
This implies that these parameters can be written as $u=\cosh(r)$ and $\upsilon=\sinh(r)$.
In this way, we can characterize the system by a single parameter, i.e., squeezing parameter $r$.
In addition, $\omega_{a}=\omega_{1}-\omega_{0}=\omega_{2}+\omega_{0}$. This means the two decay channels denoted in red of Fig.~1(b) correspond to phonon emission with the same energy $\omega_{a}$.
Hence, the generalized interaction Hamiltonian in the situation of $N$ SiV centers is
\begin{eqnarray}\label{ME14}
\hat{H}_{\text{I}}&=&\sum_{j}\alpha u(x_{j},t)(D_{j}^{+}e^{i\omega_{a}t}+\text{H.c.}).
\end{eqnarray}

We define $\rho_{S}$ as the reduced density matrix for the SiV centers, and the dynamics of the SiV centers are governed by the Born-Markovian master equation \cite{PhysRevLett.70.2269,PhysRevLett.70.2273,PhysRevLett.110.080502,PhysRevA.100.052332}
\begin{eqnarray}\label{ME15}
\frac{d}{dt}\rho_{S}=\sum_{j,m}J_{j,m}[D_{j}^{-}\rho_{S}(t)D_{m}^{+}-\rho_{S}(t)D_{m}^{+}D_{j}^{-}]+\text{H.c.},
\notag\\
\end{eqnarray}
where
\begin{eqnarray}\label{ME16}
J_{j,m}=\int_{0}^{t}dt(e^{-i\omega_{a}t}+e^{i\omega_{a}t})\sum_{n,k}g_{n,k}^{2}e^{-i\omega_{n,k}t}e^{ik(x_{j}-x_{m})}.
\notag\\
\end{eqnarray}
This equation gives the dipole-dipole coupling strength between the SiV centers at position $x_{j}$ and $x_{m}$.
In this process, we assume the phonon field is in the vacuum.
It means $\langle a_{n,k}^{\dagger}a_{n,k}\rangle=0$.

\begin{figure}
\includegraphics[width=7cm]{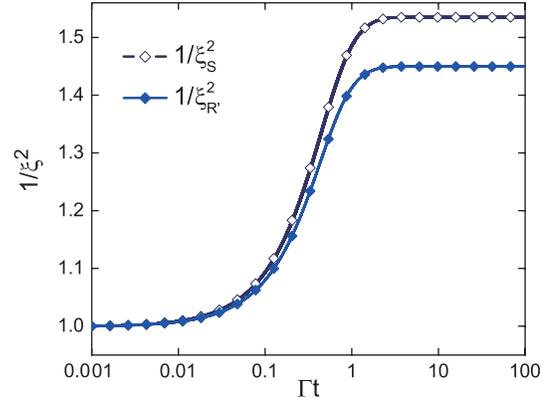}
\caption{\label{fig:3}(Color online)
  Time evolution of the two spin squeezing parameters $1/\xi_{R^{'}}^{2}$ and $1/\xi_{S}^{2}$ with $N=4$, with $r=0.2$. }
\end{figure}

In order to get the explicit expressions of $J_{j,m}$, we need some mathematical calculations.
The first step is replacing the summation of $k$ by an integral, and then transforming the $k$ integral into an energy integral by assuming the linear dispersion relationship $\omega_{n,k}=vk$ \cite{PhysRevLett.110.080502}. We finally have
\begin{eqnarray}\label{ME17}
J_{j,m}=\frac{\Gamma}{2}e^{i\omega_{a}|x_{j}-x_{m}|},
\end{eqnarray}
where $\Gamma=\alpha^{2}\gamma(\omega_{a})$ is the collective decay rate, with $\gamma(\omega)=|g_{n,k}|^{2}D(\omega)/\pi$, and $D(\omega)=(L/2\pi)|\partial_{k}\omega_{n,k}|$ the state density.
Generally, the coupling between dipoles decays with the distance. One can find this conclusion easily from $J_{j,m}$.
In this scheme, we choose the distance $x_{n}=n\lambda$ ($n\in\mathbb{Z}$) between two centers, and $\lambda=2\pi/k$, with the purpose of canceling the dipole-dipole interaction by destructive interferences \cite{PhysRevLett.110.080502,PhysRevB.82.075427,Chang_2012,PhysRevA.95.033818}.
Eventually, we get the Dicke superradiant model \cite{PhysRev.93.99} described by
\begin{eqnarray}\label{ME18}
\frac{d}{dt}\rho_{S}=\frac{\Gamma}{2}(D^{-}\rho_{S} D^{+}-D^{+}D^{-}\rho_{S})+\text{H.c.},
\end{eqnarray}
where $D^{+/-}=\sum_{j}D_{j}^{+/-}$ is the collective spin-squeezed operators.
To quantify the degree of entanglement, we introduce the spin squeezing parameter $\xi_{R^{'}}^{2}$
\begin{eqnarray}\label{ME19}
\xi_{R^{'}}^{2}=\frac{N(\Delta S_{x})^{2}}{\langle S_{y}\rangle^{2}+\langle S_{z}\rangle^{2}},
\end{eqnarray}
where  $S_{\beta}=\Sigma_{j}\sigma_{j}^{\beta}/2$, $(\beta=x,y,z)$.  We define the Pauli matrices $\sigma_{z}=|2\rangle\langle2|-|1\rangle\langle1|$, $\sigma_{+}=|2\rangle\langle1|$, and $\sigma_{-}=|1\rangle\langle2|$.
Spin squeezing is not only the measure of entanglement, but also has important applications in quantum metrology and quantum information processing \cite{PhysRevLett.110.080502,PhysRevA.47.5138,S_rensen2001,PhysRevA.50.67}.
The spin squeezing parameter $\xi_{R^{'}}^{2}$ is proposed in Ref. \cite{S_rensen2001}, and it can be viewed as a generalization of the spin squeezing parameter $\xi_{R}^{2}$  proposed by Wineland et al. in the study of spectroscopy \cite{PhysRevA.46.R6797,PhysRevA.50.67}.
When $\xi_{R^{'}}^{2}<1$, the states can be shown to be entangled.
The definitions of spin squeezing are not uniform.
Another spin squeezing definition is given by Kitagawa and Ueda \cite{PhysRevA.47.5138,wolfe2014spin}
\begin{eqnarray}\label{ME20}
\xi_{S}^{2}&=&\frac{2}{N}[\langle S_{1}^{2}+S_{2}^{2}\rangle-\sqrt{\langle S_{1}^{2}-S_{2}^{2}\rangle^{2}+\langle S_{1}S_{2}+S_{2}S_{1}\rangle^{2}}], \nonumber \\
 \end{eqnarray}
where
\begin{eqnarray}
S_{1}&=&S_{y}\cos\phi-S_{x}\sin\phi  \notag\\
S_{2}&=&S_{x}\cos\theta\cos\phi+S_{y}\cos\theta\sin\phi-S_{z}\sin\theta, \notag\\
\theta &=& \arccos(\langle S_{z}\rangle/\sqrt{\langle S_{x}\rangle^{2}+\langle S_{y}\rangle^{2}+\langle S_{z}\rangle^{2}}) \notag\\
\phi &=& \arctan(\langle S_{y}\rangle/\langle S_{x}\rangle).
\end{eqnarray}
These two spin squeezing definitions $\xi_{S}^{2}$ and $\xi_{R'}^{2}$ are not equivalent.
In Fig.~\ref{fig:3} we plot the time evolution of the spin squeezing parameters for $N=4$ to compare these two definitions.
We find that $\xi_{S}^{2} < \xi_{R^{'}}^{2}$, which is consistent with the conclusion in Ref. \cite{MA201189}.
It has been proved that the spin squeezing parameters $\xi_{S}^{2}$ and $\xi_{R^{'}}^{2}$  are closely related to pairwise or many-body entanglement \cite{PhysRevA.68.012101,S_rensen2001,MA201189}. Note that such steady-state entanglement in
Dicke superradiance has also been studied in a different context \cite{wolfe2014spin}, which  discussed
an important aspect of the entanglement witness $\xi_{S}^{2}$.
In this work, we choose $\xi_{R^{'}}^{2}$ as the criterion to detect many body entanglement, and for simplification we drop the subscribe $R^{'}$ in the following.

\begin{figure}
\includegraphics[width=7cm]{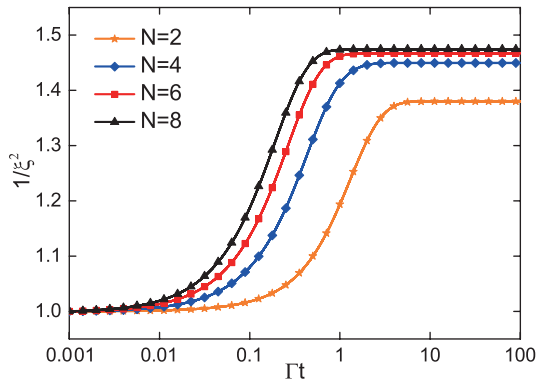}
\caption{\label{fig:4}(Color online)
  Time evolution of the squeezing parameter $1/\xi^{2}$ with $N=2,4,6,8$. }
\end{figure}

We plot in Fig.~\ref{fig:4} the time evolution of the squeezing parameter $1/\xi^{2}$.
Here, we choose the value of the parameter $r=0.2$, and $N=2,\ 4,\ 6,\ 8$.
For two SiV centers, the time for the system to reach the steady state is about $\Gamma t=10$, and the value of the parameter $1/\xi^{2}$ is about 1.4.
This means that these two centers are entangled.
When the number of SiV centers $N=4,\ 6,\ 8$, it only takes $\Gamma t<1$ to realize the steady state entanglement, and $1/\xi^{2}$ is about 1.5.
The numerical result indicates that the degree of entanglement can be enhanced with the increasing of the number of SiV centers.
Moreover, it takes less time for a large $N$ to reach the steady state. Therefore, it will be more efficient to prepare the targeted states with a high
fidelity with a large number of SiV centers.

\begin{figure}[t]
\includegraphics[width=7cm]{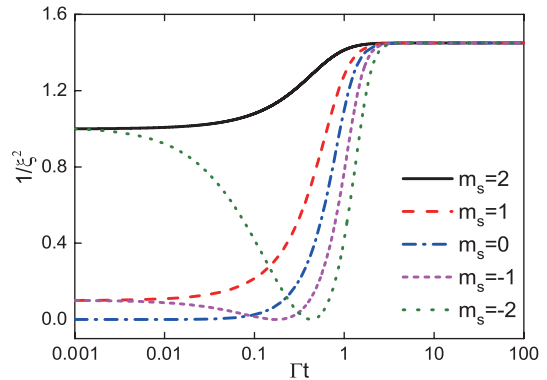}
\caption{\label{fig:5}(Color online)
  Time evolution of the squeezing parameter $1/\xi^{2}$ for different initial states $|N/2,m_{s}\rangle$. Here the parameters are $N=4$, and $r=0.2$. Note that the steady state does not depend on the initial state in the subspace $S=N/2$. }
\end{figure}

We note that the Hamiltonian commutes with the total spin $\vec{S}^{2}$.
This means, if the total spin of the system is given, the system will eventually reach a unique steady state regardless of the initial spin projection \cite{PhysRevLett.110.120402}.
We assume that the initial state is $|\Psi_{0}\rangle=|S,m_{s}\rangle=|N/2,N/2\rangle$, and the system evolves within the sector $S=N/2$.
We next check the time evolution of the squeezing parameter $1/\xi^{2}$ for different initial states in Fig.~\ref{fig:5}.
One can find that all the squeezing parameters evolve asymptotically to the same stable value.
This result agrees with what we discuss above.

\begin{figure}[b]
\includegraphics[width=7cm]{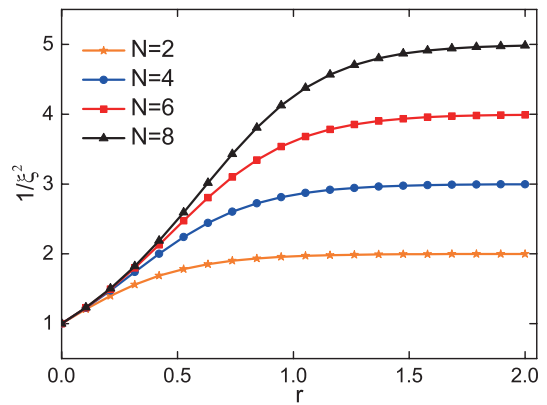}
\caption{\label{fig:6}(Color online)
  Spin squeezing $1/\xi^{2}$ as a function of the squeezing parameter $r$ at steady state, with different numbers of SiV centers ($N=2,\ 4,\ 6,\ 8$) . }
\end{figure}

Finally, In Fig.~\ref{fig:6} we plot the calculation of the spin squeezing in the steady state as a function of the squeezing parameter $r$.
When $N=2$, the maximum value of $1/\xi^{2}$ rises to 2 when increasing the parameter $r$, and for $N=8$, the maximum value of $1/\xi^{2}$ can reach 5. From the figure, one can find an enhancement of the maximum value of the spin squeezing when  increasing  the number of the SiV centers.

For the realistic experimental condition in our scheme, we should take the dephasing rate $\Gamma_{D}$ into our consideration, and add this inevitable adverse factor into our numerical simulation.
According to Eq.~(\ref{ME18}), we can obtain
\begin{eqnarray}\label{ME21}
\frac{d}{dt}\rho_{S}=\Gamma\mathcal{D}[D^{-}]\rho_{S}+\Gamma_{D}\sum_{j}\mathcal{D}[\sigma_{z}^{j}]\rho_{S},
\end{eqnarray}
where $\mathcal{D}[O]\rho=O\rho O^{\dag}-1/2(O^{\dag}O\rho+\rho O^{\dag}O)$, and $\Gamma_{D}=1/2T_{2}$, with $T_{2}^{-1}$  the single spin dephasing rate \cite{PhysRevB.100.214103,PhysRevA.100.043825,PhysRevLett.124.053601}.

\begin{figure}[t]
\includegraphics[width=8.5cm]{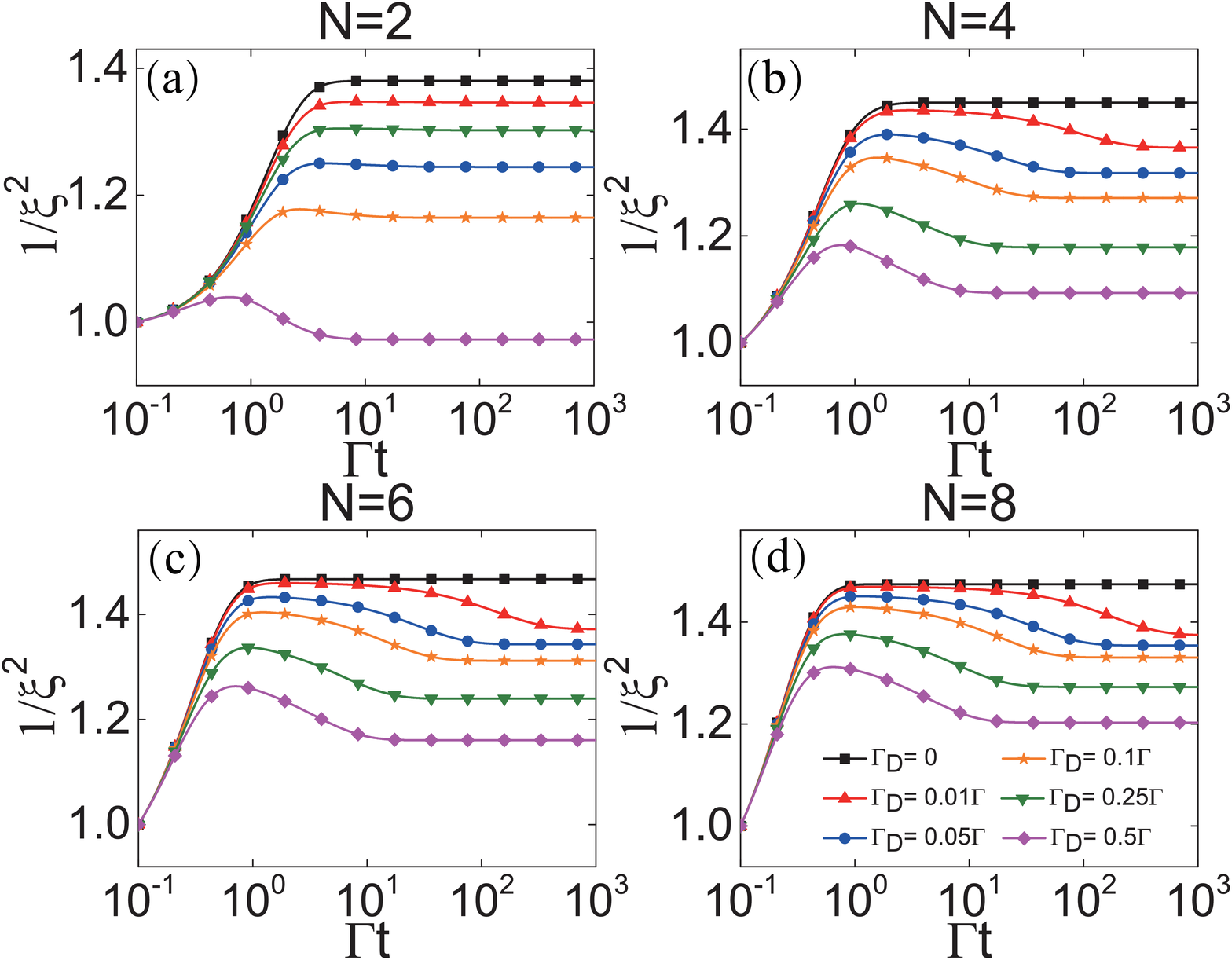}
\caption{\label{fig:7}(Color online)
Time evolution of the squeezing parameter $1/\xi^{2}$  in the presence of dephasing.}
\end{figure}

\begin{figure}[b]
\includegraphics[width=8.5cm]{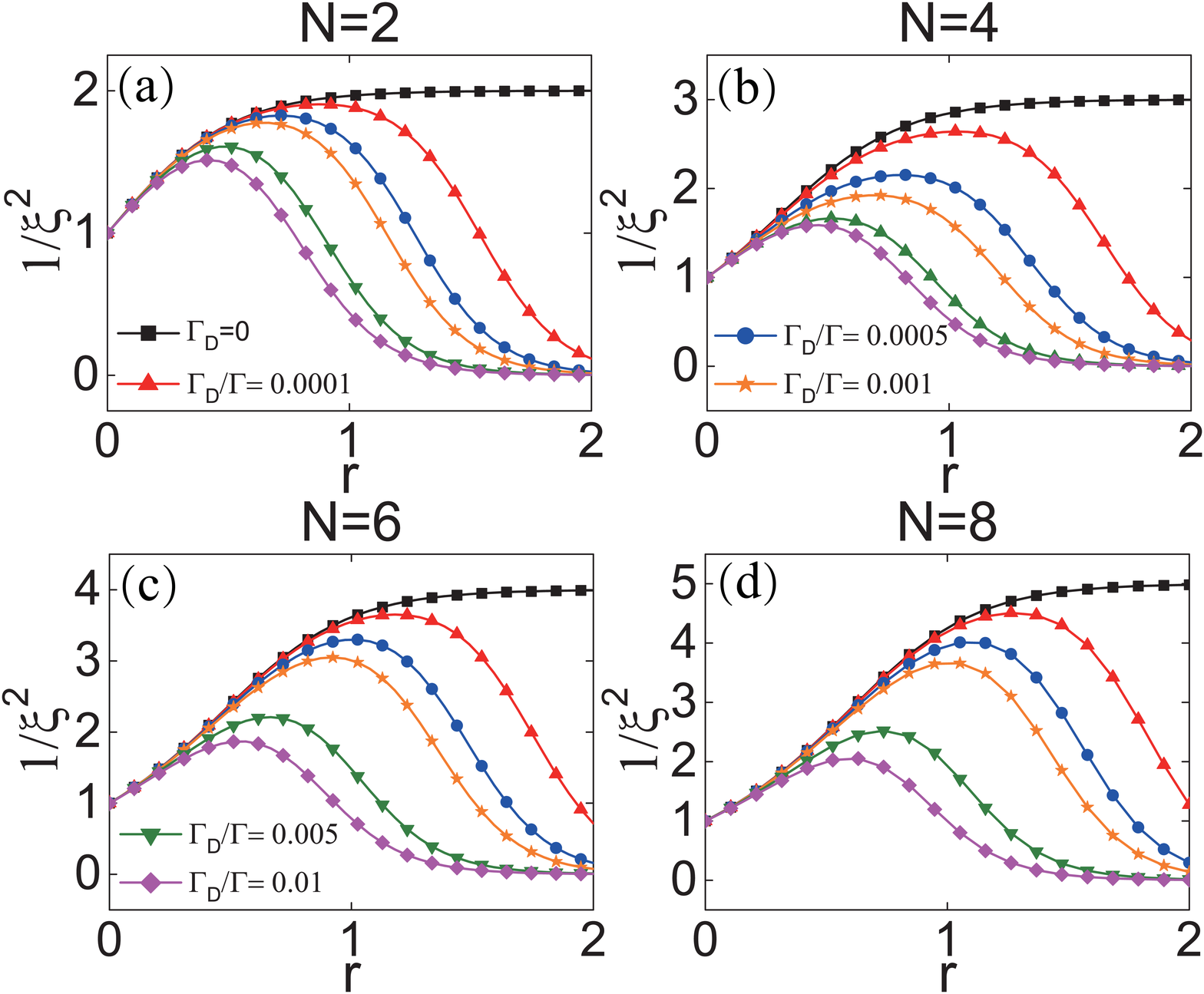}
\caption{\label{fig:8}(Color online)
The degree of entanglement as a function of the squeezing parameter $r$  in the presence of dephasing.}
\end{figure}

We plot the time evolution of the spin squeezing parameter for $N=2,\ 4,\ 6,\ 8$ spins in Fig.~\ref{fig:7}.
Here, we still choose the value of the squeezing parameter as $r=0.2$.
We can obtain the maximal entanglement in the case of $\Gamma_{D}=0$,
which has been discussed above.
For $\Gamma_{D}\neq 0$, the spin dephasing affects the steady state entanglement, and
the degree of entanglement will be reduced. As expected, this adverse effect will be significant as we increase $\Gamma_{D}$.
In addition, the time interval for the system to reach the steady state entanglement is negatively related to $\Gamma_{D}$.
As illustrated in Fig.~\ref{fig:7}, for a large $N$, as long as we take a much larger dephasing rate,
the time interval for reaching the steady entanglement will be shorter during this dynamical evolution process.
In Fig.~\ref{fig:8}, we present the calculation of the spin squeezing in the steady state as a function of the squeezing parameter $r$ for $N=2,\ 4,\ 6,\ 8$.
The dephasing effect competes with the collective decay, which results in an optimal $r$ to generate the maximal steady state entanglement. And this optimal spin parameter $r$ is inversely related to the dephasing rate $\Gamma_{D}$.

\section{The feasibility of this scheme}
In this work, we propose a spin-phononic waveguide system, where separated SiV centers are coupled by quantized phonon modes of a one-dimensional diamond waveguide. Based on state-of-the-art nanofabrication techniques, several experiments
have demonstrated the generation of SiV center arrays
through ion implantation techniques \cite{PhysRevApplied.7.064021,doi:10.1021/nl102066q}.
For instance, three arrays of SiV centers are integrated into the diamond cantilever, and are placed at determined position to control the interaction of its spins with the thermal bath in Ref. \cite{Sohn2018}.
In general, the proposed setup can be implemented experimentally where an array of SiV centers are embedded in a 1D diamond waveguide.

For the phonon diamond waveguide, the length and cross section are $L=100$ $\mu$m and $A=100 \times 100$ $\text{nm}^{2}$.
In addition, the material properties of diamond waveguide are $\rho=3500$ $\text{kg/m}^{3}$, $E=1050$ GPa and $\nu=0.2$.
In this case, the group velocity along the waveguide is $v=1.71\times10^{4}$ $\text{m/s}$, and the single SiV center can couple to the phonon mode with a strength $g/2\pi=16$ MHz.

For the SiV center in diamond, the ground-state splitting is about $\Delta/2\pi=46$ GHz.
And the transitions between $|3\rangle\leftrightarrow|2\rangle$ and $|4\rangle\leftrightarrow|1\rangle$ can be implemented by using a microwave field \cite{PhysRevLett.120.213603} or via an equivalent optical Raman process \cite{PhysRevLett.120.213603}, which has already been carried out experimentally.
In our model, the precise location of the SiV centers is required with the purpose of cancelling the dipole-dipole interactions by destructive interference.
The diamond waveguide permits us to place the SiV centers permanently at the position $x_{n}=n\lambda$ \cite{PhysRevLett.110.080502}, where $\lambda\approx200$ nm is the phonon wavelength.
Under this consideration, a waveguide with $L=100$ $\mu$m can contain 500 SiV centers, approximately.

At 100 mK temperature, the spin dephasing time of a single SiV center is about $\gamma_{s}/2\pi=100$ Hz.  Thus the spin coherence time is $T_{2}\sim10$ ms.
Assuming $\Omega_{1,2}/2\pi\sim10$ MHz and $\Delta_{1,2}/2\pi\sim100$ MHz, the effective decay rate is $\Gamma/2\pi=160$ kHz.
According to Fig.~4, the time for the system to reach the steady state entanglement is about $\Gamma t\lesssim10$, i.e., $T\thicksim10$ $\mu$s, which is much shorter than the spin coherence time $T_{2}$. This shows that the condition for the steady state entanglement can be reached.

\section{Conclusion}
In conclusion, we have proposed a scheme to generate multiparticle entanglement of solid-state qubits by embedding an array of SiV centers in a one dimensional phononic diamond waveguide.
We study the efficient coupling between the SiV centers and the phononic waveguide modes.
The unique structure of the waveguide allows us to switched off the dipole-dipole interactions via destructive interferences by appropriately choosing the distance between nearby SiV centers.
We can achieve the Dicke supperradiant model under appropriate external driving fields.
We also discuss the degree of entanglement in the presence of realistic decoherence.
This scheme may provide a realistic and feasible platform for quantum information processing  with spins and phonons in a solid-state system.
\section*{Acknowledgments}
This work is supported by the NSFC under Grant No. 11774285 and the Fundamental Research Funds for the Central Universities.

\begin{appendix}

\section{The Hamiltonian of SiV centers}

The detailed derivation for the Hamiltonian of SiV centers has been discussed in Refs. \cite{PhysRevLett.112.036405,PhysRevB.94.214115,PhysRevLett.120.213603}. Here we follow their discussions and present the key results.
We start from the  Hamiltonian (Eq.~(\ref{ME2}) in main text), which is in the basis spanned by the degenerate eigenstates $|e_{x},\uparrow\rangle$, $|e_{x},\downarrow\rangle$, $|e_{y},\uparrow\rangle$ and $|e_{y},\downarrow\rangle$
\begin{eqnarray}\label{MEA1}
\hat{H}_\text{SiV}&=&-\lambda_{g}\begin{bmatrix} 0 &i \\ -i &0 \end {bmatrix} \otimes \frac{1}{2} \begin{bmatrix} 1&0 \\ 0&-1 \end {bmatrix}+\begin{bmatrix} \Upsilon_{x}&\Upsilon_{y} \\ \Upsilon_{y}&-\Upsilon_{x} \end {bmatrix} \otimes \mathbb{I} \notag\\
&+&f\gamma_{L} \begin{bmatrix} 0 &i \\ -i &0 \end {bmatrix} \otimes B_{0}\mathbb{I}+\gamma_{S}B_{0} \mathbb{I}  \otimes \frac{1}{2}\begin{bmatrix} 1 &0 \\ 0 & -1 \end {bmatrix}.
\end{eqnarray}

After some simplifications and neglecting the effect of the reduced orbital Zeeman interaction, we obtain the total Hamiltonian of SiV centers
\begin{widetext}
\begin{eqnarray}\label{MEA2}
\hat{H}_\text{SiV}=\begin{bmatrix} \frac{1}{2}\gamma_{S}B_{0}+\Upsilon_{x} & 0 & -\frac{i}{2}\lambda_{g}+\Upsilon_{y} & 0 \\ 0 & -\frac{1}{2}\gamma_{S}B_{0}+\Upsilon_{x} & 0 & \frac{i}{2}\lambda_{g}+\Upsilon_{y} \\ \frac{i}{2}\lambda_{g}+\Upsilon_{y} & 0 & \frac{1}{2}\gamma_{S}B_{0}-\Upsilon_{x} & 0 \\ 0 & -\frac{i}{2}\lambda_{g}+\Upsilon_{y} & 0 & -\frac{1}{2}\gamma_{S}B_{0}-\Upsilon_{x}  \end {bmatrix}.
\end{eqnarray}
\end{widetext}
Diagonalizing the above matrix, we have the eigenenergies
\begin{eqnarray}\label{MEA3}
E_{3,1}&=&-\frac{1}{2}\gamma_{S}B_{0}\pm\sqrt{\Upsilon^{2}+\frac{1}{4}\lambda_{g}^{2}},  \notag\\
E_{4,2}&=&\frac{1}{2}\gamma_{S}B_{0}\pm\sqrt{\Upsilon^{2}+\frac{1}{4}\lambda_{g}^{2}}.
\end{eqnarray}
Here, $\Upsilon=\sqrt{\Upsilon_{x}^{2}+\Upsilon_{y}^{2}}$ is the Jahn-Teller coupling strength.
It has been shown that the spin-orbit interaction and the Jahn-Teller effect both lift the orbital degeneracy of ground states of SiV centers \cite{PhysRevLett.112.036405}.
Besides, the spin-orbit coupling strength $\lambda_{g}\gg \Upsilon_{x,y}$. We thus neglect the Jahn-Teller effect on the orbital states.
In this case, the corresponding normalized eigenstates are
\begin{eqnarray}\label{MEA4}
|1\rangle &\approx& (|e_{x},\downarrow\rangle - i|e_{y},\downarrow\rangle)/\sqrt{2}=|e_{-},\downarrow\rangle, \notag\\
|2\rangle &\approx& (|e_{x},\uparrow\rangle + i|e_{y},\uparrow\rangle)/\sqrt{2}=|e_{+},\uparrow\rangle, \notag\\
|3\rangle &\approx& (|e_{x},\downarrow\rangle + i|e_{y},\downarrow\rangle)/\sqrt{2}=|e_{+},\downarrow\rangle, \notag\\
|4\rangle &\approx& (|e_{x},\uparrow\rangle - i|e_{y},\uparrow\rangle)/\sqrt{2}=|e_{-},\uparrow\rangle.
\end{eqnarray}
After shifting the whole energy level of SiV centers, i.e., choosing $E_{1}=0$, we then have the Hamiltonian form  Eq.~(\ref{ME3}) in the main text.

\section{The quantization of the phonon waveguide modes}
We follow the discussion in Ref. \cite{PhysRevLett.120.213603} to obtain the key results  on the quantization of the phonon waveguide modes.
We consider a 1D diamond waveguide with a cross section $A$ and a length $L\gg A$.
The phonon modes can be viewed as  elastic waves in the elastic theory, characterized by the displacement field $\vec{u}(\vec{r},t)$ \cite{PhysRevLett.120.213603}.
For a linear isotropic medium, the displacement field is described by the equation of motion
\begin{eqnarray}\label{MEB1}
\rho\frac{\partial^{2}}{\partial t^{2}}\vec{u}=(\lambda+\mu)\vec{\nabla}(\vec{\nabla}\cdot\vec{u})+\mu\vec{\nabla}^{2}\vec{u},
\end{eqnarray}
where $\rho=3500$ $\text{kg}/ \text{m}^{3}$ is the mass density of diamond waveguide, $\lambda$ and $\mu$ are the Lam\'{e} constants
\begin{eqnarray}\label{MEB2}
\lambda=\frac{\nu E}{(1+\nu)(1-2\nu)},\mu=\frac{E}{2(1+\nu)}.
\end{eqnarray}
For diamond, we use the Young's modulus $E=1050$ GPa and Poisson ratio $\nu=0.2$.

Under the periodic boundary conditions $k=2\pi m/L$ ($m\in\mathbb{Z}$), the solution of Eq.~(\ref{MEB1}) is given by
\begin{eqnarray}\label{MEB3}
\vec{u}(\vec{r},t)=\frac{1}{\sqrt{2}}\sum_{n,k}\vec{u}_{n,k}^{\perp}(y,z)[A_{n,k}(t)e^{ikx}+\text{c.c.}].
\end{eqnarray}
Here, amplitudes $A_{n,k}(t)$ satisfy the oscillating equation $\ddot{A}_{n,k}(t)+\omega_{n,k}^{2}A_{n,k}(t)=0$ \cite{PhysRevLett.120.213603}.
The value of the  mode frequencies $\omega_{n,k}$ and the transverse mode profile $\vec{u}_{n,k}^{\perp}(y,z)$ can  be obtained from  numerical solutions, and the $\vec{u}_{n,k}^{\perp}(y,z)$ are orthogonal and normalized to
\begin{eqnarray}\label{MEB4}
\frac{1}{A}\int dydz\vec{u}_{n,k}^{\perp}(y,z)\cdot\vec{u}_{m,k}^{\perp}(y,z)=\delta_{nm}.
\end{eqnarray}
The quantization of displacement field is similar to the electromagnetic field in quantum optics.
Taking the equivalence
\begin{eqnarray}\label{MEB5}
A_{n,k} \rightarrow \sqrt{\frac{\hbar}{\rho V\omega_{n,k}}}a_{n,k}^{\dag},  A_{n,-k}^{\ast} \rightarrow \sqrt{\frac{\hbar}{\rho V\omega_{n,k}}}a_{n,-k},
\end{eqnarray}
we can obtain the quantized displacement field \cite{PhysRevLett.120.213603}
\begin{eqnarray}\label{MEB6}
\vec{u}(\vec{r})=\sum_{n,k}\sqrt{\frac{\hbar}{2\rho V\omega_{n,k}}}\vec{u}_{n,k}^{\perp}(y,z)(a_{n,k}e^{ikx}+a_{n,k}^{\dag}e^{-ikx}),
\notag\\
\end{eqnarray}
and the quantized phonon modes
\begin{eqnarray}\label{MEB7}
\hat{H}_{\text{ph}}=\sum_{n,k}\hbar\omega_{n,k}a_{n,k}^{\dag}a_{n,k}.
\end{eqnarray}

\section{Strain coupling}
We follow the discussion in Ref. \cite{PhysRevLett.120.213603} to obtain the key results  on strain coupling between SiV centers and phononic waveguide modes.
For small displacements and in the Born-Oppenheimer approximation, the strain coupling within the framework of linear elasticity theory can be given by
\begin{eqnarray}\label{MEC1}
\hat{H}_{\text{strain}}=\sum_{ij}V_{ij}\epsilon_{ij}.
\end{eqnarray}
Here, $i$,$j$ label the coordinate axes.
$V_{ij}$ is an operator acting on the electronic states of the SiV defect and $\epsilon_{ij}$ is the strain field tensor \cite{Maze_2011}.
The local strain tensor is defined as
\begin{eqnarray}\label{MEC2}
\epsilon_{ij}=\frac{1}{2}(\frac{\partial u_{i}}{\partial x_{j}}+\frac{\partial u_{j}}{\partial x_{i}}),
\end{eqnarray}
with $u_x$ ($u_y$, $u_z$) representing the quantized displacement field along $x$ ($y$, $z$) at position of the SiV center.
We assume the axes as shown in Fig.~\ref{fig:1}(a).
By projecting the strain tensor onto the irreducible representation of $D_{3d}$, the Hamiltonian can be rewritten in terms of the electronic states of the SiV defect
\begin{eqnarray}\label{MEC3}
\hat{H}_{\text{strain}}=\sum_{r}V_{r}\epsilon_{r},
\end{eqnarray}
where $r$ runs over the irreducible representations.
It can be shown that the only contributing representations are the one-dimensional representation $A_{1g}$ and the two-dimensional representation $E_{g}$.
Due to the inversion symmetry, the ground states of the SiV center transform as $E_{g}$, and the excited states transform as $E_{u}$.
Therefore, one can find that strain can couple independently to orbital within the ground and excited manifolds.
Limiting only to the ground state, the terms in Eq.~(\ref{MEC3}) are given by
\begin{eqnarray}\label{MEC4}
\epsilon_{A_{1g}}&=&t_{\bot}(\epsilon_{xx}+\epsilon_{yy})+t_{\|}\epsilon_{zz}, \notag\\
\epsilon_{E_{gx}}&=&d(\epsilon_{xx}-\epsilon_{yy})+f\epsilon_{zx}, \notag\\
\epsilon_{E_{gy}}&=&-2d\epsilon_{xy}+f\epsilon_{yz},
\end{eqnarray}
where $t_{\bot}$, $t_{\|}$, $d$, $f$ are four strain-susceptibility parameters.
Furthermore, the effects of these strain components on the electronic states are given by
\begin{eqnarray}\label{MEC5}
V_{A_{1g}}&=&|e_{x}\rangle\langle e_{x}|+|e_{y}\rangle\langle e_{y}|, \notag\\
V_{E_{gx}}&=&|e_{x}\rangle\langle e_{x}|-|e_{y}\rangle\langle e_{y}|, \notag\\
V_{E_{gy}}&=&|e_{x}\rangle\langle e_{y}|+|e_{y}\rangle\langle e_{x}|.
\end{eqnarray}

As the energy shift of all ground states induced by symmetry local distortions is equal, it can be neglected.
Finally, we write the strain Hamiltonian using the basis spanned by the eigenstates of the spin-orbit coupling $|e_{+}\rangle$, and $|e_{-}\rangle$
\begin{eqnarray}\label{MEC6}
\hat{H}_{\text{strain}}=\epsilon_{E_{gx}}(\hat{L}_{-}+\hat{L}_{+})-i\epsilon_{E_{gy}}(\hat{L}_{-}-\hat{L}_{+}),
\end{eqnarray}
where $\hat{L}_{+}=\hat{L}_{-}^{\dag}=|3\rangle\langle1|+|2\rangle\langle4|$.
By decomposing the local displacement field in terms of the vibrational eigenmodes, the resulting strain coupling can be written as
\begin{eqnarray}\label{MEC7}
\hat{H}_{\text{strain}}=\sum_{n,k}[g_{n,k}(J_{+}+J_{-})a_{n,k}e^{ikx}+\text{H.c.}],
\end{eqnarray}
with the coupling strength having the form
\begin{eqnarray}\label{MEC8}
g_{n,k}&=&g_{0}\frac{1}{|k|}[(ik\vec{u}_{n,k}^{\perp,x}+ik\frac{f}{d}\frac{\vec{u}_{n,k}^{\perp,z}}{2}+\frac{f}{d}\frac{\partial_{z}\vec{u}_{n,k}^{\perp,x}}{2}-\partial_{y}\vec{u}_{n,k}^{\perp,y})\notag\\
&-&i(ik\vec{u}_{n,k}^{\perp,y}+\frac{f}{d}\frac{\partial_{y}\vec{u}_{n,k}^{\perp,z}}{2}+\frac{\partial_{z}\vec{u}_{n,k}^{\perp,y}}{2}+\partial_{y}\vec{u}_{n,k}^{\perp,x})],
\end{eqnarray}
where $g_{0}=\sqrt{\hbar k^{2}d^{2}/2\rho V\omega_{n,k}}$.

\end{appendix}
%\bibliography{ref}

\end{document}